\documentclass[12pt, twoside]{amsart}
\usepackage{inputenc}

\scrollmode 
\usepackage{amsmath}
\usepackage{amsthm}
\usepackage{amssymb}
\usepackage{amsfonts}
\usepackage[matrix,arrow,curve]{xy}
\newtheorem*{thm}{Theorem}
\newtheorem{theorem}{Theorem}[section]
\newtheorem{lemma}[theorem]{Lemma}
\newtheorem{proposition}[theorem]{Proposition}

\theoremstyle{definition}

\newtheorem{remark}[theorem]{Remark}

\newcommand{\sign}{\mathop{\rm sign}}
\newcommand{\set}[1]{\left\{#1\right\}}

\newcommand{\Mat}{\mathop{\mathrm{Mat}}}

\newcommand{\nc}{\newcommand}
\nc{\Symm}{{\on{Sym}}}

\newcommand{\on}{\operatorname}

\nc{\cE}{{\cal E}}

\renewcommand{\d}{{\mathfrak d}}
\nc{\SL}{{\mathfrak sl}}

\nc{\HH}{{\mathfrak h}}

\newcommand{\q}{{\mathfrak{q}}}

\nc{\wh}{\widehat}\nc{\wt}{\widetilde}

\newcommand{\half}{\textstyle{\frac{1}{2}}}

\newcommand{\ben}{\begin{enumerate}}
\newcommand{\een}{\end{enumerate}}
\newcommand{\ad}{{\text{ad}}}

\newcommand{\CC}{{\mathbb{C}}}

\newcommand{\RR}{{\mathbb R}}

\newcommand{\ZZ}{{\mathbb{Z}}}

\newcommand{\cC}{{\mathcal C}}

\newcommand{\cM}{{\mathcal M}}
\newcommand{\cW}{{\mathcal W}}
\newcommand{\cV}{{\mathcal V}}

\newcommand{\cH}{{\mathcal H}}

\newcommand{\diag}{\mathrm{diag}\,}
\renewcommand{\d}{\partial}

\renewcommand{\q}{\quad}

 \usepackage{hyperref}

\begin{document}

\title[Poisson groups and   Schroedinger equation on the circle]
{Poisson groups and  differential Galois theory of Schroedinger
equation on the circle}

\author{Ian Marshall}
\address{Mathematics Department, University of Loughborough, UK}
\email{ian.marshall@math.unige.ch}

\author{Michael Semenov-Tian-Shansky}
\address{Institut Math\'ematique de Bourgogne, Dijon, France and Steklov Mathematical Institute, St.~Petersburg, Russia
}
\email{semenov@u-bourgogne.fr}

\maketitle

\begin{abstract}
We combine the projective geometry approach to Schroedinger
equations on the circle and  differential Galois theory with the
theory of Poisson Lie groups to construct a natural Poisson
structure on the space of wave functions (at the  zero energy
level). Applications to KdV-like nonlinear equations are discussed.
The same approach is applied to 2$^\text{nd}$ order difference
operators on a one-dimensional lattice, yielding an extension of the
lattice Poisson Virasoro algebra.
\end{abstract}

\section*{Introduction}

It is well known that the space $\mathcal{H}$ of Schroedinger
operators on the circle
$$
H=-\d_x^2 - u,\qquad u\in C^\infty(S^1), \quad S^1\simeq
\mathbb{R}/2\pi \mathbb{Z},
$$
may be regarded as the phase space for the KdV hierarchy (with
periodic boundary conditions). It carries a family of natural
Poisson structures which play an important r\^ole in the Hamiltonian
description of the KdV flows. In this letter we shall be concerned
with the so called second Poisson structure for the KdV equation
associated with the third order differential operator
\begin{equation}\label{1}
  l=\tfrac12
  \d_x^3+u\d_x+\d_x u.
\end{equation}
 This Poisson structure may be regarded as
the Lie--Poisson bracket associated with the Virasoro algebra and
arises as a result of the identification of $\mathcal{H}$ with (a
hyperplane in) the dual space of the Virasoro algebra. Our aim
 is to describe its extension to the space of \emph{wave
functions}, i.e., of
  solutions of the Schroedinger equations (at  zero energy level). Despite its apparent simplicity, this question involves
several nontrivial points and has not been fully explored in the
existing literature.\footnote{We do not discuss the generalization
to the case of higher order differential operators, as well as
relation to the Drinfeld--Sokolov theory \cite{ds}. These questions
will be addressed in a separate publication.}

According to   elementary theory, for a given $u$ the space $V=V_u$
of solutions of the Schroedinger equation
\begin{equation}\label{hill}
 -\psi''-u\psi=0
\end{equation}
is 2-dimensional and for any two solutions $\phi$, $\psi$ their
wronskian $W=\phi\psi'-\phi'\psi$ is constant. An element $w\in V$
may be  regarded as a non-degenerate quasi-periodic plane curve (the
non-degeneracy condition means that $w\wedge w'$ is nowhere zero).
 There exists a matrix $M\in SL(2)$ (the
monodromy matrix) such that, writing
elements of $V$ as row vectors $w=(\phi,\psi)$,
\[
w(x+2\pi n)=w(x)M^n, \q n\in \mathbb{Z}.
\]
The group $G=SL(2)$ acts naturally on $V$ (preserving the wronskian)
by right multiplication.   $G$ plays a key r\^ole in the geometry of
 $\cH$ in its double guise of the
differential Galois group of equation \eqref{hill} and of the group
of projective transformations. Both aspects are completely
classical; the
novel element introduced in the present paper consists in their interaction
with the Poisson geometry.

Let us   recall how the Schroedinger equation is seen from the
viewpoint of
  projective geometry. The following assertion is well known
(see \cite{ot}).
\begin{thm}
(i) Any pair of linearly independent solutions of the Schroedinger
equation defines a  non-degenerate quasi-periodic projective curve
$\gamma:\mathbb{R}\to \mathbb{C}P_1$ such that
$\gamma(x+2\pi)=\gamma(x)M$. Any two projective curves associated
with a given   Schroedinger equation are related by a global
projective transformation. (ii) Conversely, any non-degenerate
quasi-periodic projective curve may be lifted to a non-degenerate
curve in $\mathbb{C}^2$ such that its wronskian is equal to 1.
\end{thm}


 In more abstract language,  $\mathcal{H}$ is the space
of projective connections on the circle. For a given $H_u=-\d^2-u\in
\mathcal{H}$ there is a natural projective line bundle
$\mathcal{P}_u\to S^1$;  the quasi-periodic projective curve
referred to above is its covariantly constant section, and the group
$G=SL(2)$, or, more precisely, the associated projective group
$PSL(2)=SL(2)/\set{\pm 1}$), its  structure group.

Without restricting the generality we may fix an affine coordinate
on $\mathbb{C}P_1$ in such a way that $\infty$ corresponds  to the
zeros of the second coordinate $\psi$ of the point on the plane
curve; with this choice $\gamma$ is replaced with the affine curve
$x\mapsto\eta(x)=\phi(x)/\psi(x)$. The potential $u$ may be restored
from $\eta$ by the formula
$$
u= \tfrac12S(\eta),
$$
where $S$ is the Schwarzian derivative
$$
S(\eta)= {\frac{\eta'''}{\eta'}} -
\tfrac{3}{2}\left({\frac{\eta''}{\eta'}}\right)^2,
$$
which has the crucial property of being   invariant under projective
transformations
$$
\eta\mapsto\frac{a\eta+c}{b\eta+d}
$$
induced by the right action of $G$.

The space $\cV$ of all quasi-periodic plane curves with wronskian 1,
or the equivalent space of projective curves (together with the
associated monodromy matrices) encodes all information about
Schroedinger
 operators. In   \cite{wilson1} G.~Wilson
 considered the extension of the KdV hierarchy to this space. To put
it in a more formal way let us note that the natural ``algebra of
observables'' associated with the KdV equation consists of local
functionals of the form
$$
F[u]=\int_0^{2\pi}F(u,\d_x u,\d^2_x u,\dots)\,dx,
$$
where $F$ is a polynomial (or, more generally, a rational) function
of $u$ and of its derivatives. We can identify the observable $F[u]$
and the corresponding density; in other words, our basic algebra of
observables is identified with the differential field
$\mathbb{C}\langle u\rangle$. In the same way, we can associate with
 the
space of solutions of the Schroedinger equation a bigger
differential field $\mathbb{C}\langle\phi, \psi\rangle$. Clearly,
$\mathbb{C}\langle\phi, \psi\rangle\supset\mathbb{C}\langle
u\rangle$; as a matter of fact, $\mathbb{C}\langle u\rangle$ is
isomorphic to  the differential subfield of $G$-invariants and hence
$\mathbb{C}\langle\phi, \psi\rangle\supset\mathbb{C}\langle
u\rangle$ is a differential Galois extension with   differential
Galois group $G=SL(2)$ (we shall speak  below simply of Galois
groups and Galois extensions, for short). Various subgroups of $G$
give rise to intermediate differential fields. In particular, for
$Z=\set{\pm I}$ the associated subfield of invariants is naturally
isomorphic to $\mathbb{C}\langle \eta\rangle$; since $Z$ is the
center of $G$, the extension $\mathbb{C}\langle \eta\rangle\supset
\mathbb{C}\langle u\rangle$ is again a Galois extension with the
Galois group $PSL(2)=SL(2)/Z$.\footnote{The Galois theory point of
view was implicit in the old paper of Drinfeld and Sokolov
\cite{ds1}, where
  wave functions for \emph{different} values of energy are
  considered, leading to  to an extended class of ``equations of KdV
  type''.
Generically, the  associated Galois group becomes in this case the
product of several copies of $SL(2)$. }

Let $B$ the subgroup of lower triangular matrices; its field of
invariants $\mathbb{C}\langle\phi, \psi\rangle^B$ may be identified
with $\mathbb{C}\langle v\rangle$, where
$v=\frac12\frac{\eta''}{\eta'}$. One has $u=v'-v^2$, which is the
classical Miura transform. Note that since $B$ is not normal in $G$,
$\mathbb{C}\langle v\rangle\supset \mathbb{C}\langle u\rangle$ is
not a Galois extension, and hence, as noted by Wilson
\cite{wilson1}, the   treatment of the Miura transform requires the
introduction of the `universal covering' algebra
$\mathbb{C}\langle\phi, \psi\rangle$.

The natural idea explored in \cite{wilson1} is the possibility to
lift the KdV flows originally defined on $\mathcal{H}$ to the bigger
space $\cV$. An important ingredient of such an extension is to
equip $\cV$ with a Poisson structure or its substitute.
  Wilson's point of view is
to look at the symplectic form, because it may be naturally pulled
back (at the expense of becoming degenerate, see \cite{wilson1}). A
closer look at the situation reveals yet another difficulty: the
relevant `variational' 2-form is an integral of a density whose
differential is not identically zero; rather  it is a closed form on
the circle and hence its contribution disappears only if we may
discard `total derivatives'. This convention, adopted in formal
variational calculus, greatly simplifies many formulae, but
sometimes hides important ``obstruction terms''. In Wilson's paper
this difficulty is avoided by the tacit assumption that the
monodromy matrix is equal to 1. Without this assumption the
degenerate 2-forms discussed in his paper are not closed; hence
finally his approach is intrinsically close to the quasi-Hamiltonian
formalism of Alekseev, Malkin and Meinrenken \cite{alexeev}. An
alternative approach,   followed in the present paper, is to look at
the Poisson structure. Of course, Poisson brackets cannot be pulled
back, and hence we have to \emph{guess} a Poisson structure on the
extended algebra and then check its consistency with the original
bracket.  Our strategy is based on the projective point of view
 outlined above. Although the space of projective curves is
our main object, it is
    natural to   start with the much bigger
space $\cW$ of all quasi-periodic plane curves,
$$
\cW=\set{(w =(\phi, \psi), M)\;|\; w(x+2\pi)=w(x)M}.
$$
The space $\cW$ contains the set $\cW'$ of all non-degenerate plane
 curves with non-zero
wronskian  as an open subset. Let
$\cC:=C^\infty(\RR/2\pi\ZZ,\mathbb{C}^\times)$ be the \emph{scaling
group} which acts on $\cW$ via
\begin{equation}\label{actionc}
f\cdot(w\,,M)=(fw\,,M).
\end{equation}
 Clearly,   $\cC$ acts freely on $\cW'$
 and the quotient may be identified with $\cV$. The action of the linear group $G=SL(2)$ on $\cW$ is via
$g\colon w\mapsto w\cdot g, \; M\mapsto g^{-1}Mg$.
 The  key condition which we use to
restrict  the choice of the Poisson structure on $\cW$ is its
\emph{covariance} with respect to the group action. This condition
puts us in the framework of Poisson group theory, as it allows  both
$\cC$ and $G$ to carry   nontrivial Poisson structures, although it
does not presume any \emph{a priori} choice of these structures. As
it
 happens, the
covariance condition together with the natural
 constraint on the wronskian make their choice
  almost completely canonical. (In particular, the
Poisson bracket on $G$ is fixed up to scaling and conjugation; it is
of the standard ``quastriangular'' type and  the case of zero
bracket is excluded.) Let us note that the Poisson structure on
$\cW$   constructed in this way is   closely related to the so
called
  \emph{exchange algebras} discovered in the end of 1980s
\cite{Babelon1}. The point of view adopted in the present paper
provides a useful and nontrivial complement to these old results in
making explicit  the hidden Poisson group aspects of
 differential Galois theory.
 It provides a natural route to the usual Virasoro algebra and also to its discrete
 analogue as discussed in \cite{ft}, \cite{v}, \cite{Babelon2} and \cite{frs}.

\section{A review of Poisson Lie groups}

Let $G$ be a Lie group with Lie algebra $\mathfrak{g}$. A Poisson
structure on $G$  is called multiplicative  if the multiplication
\[
m\colon G\times G\to G
\]
is a Poisson mapping.  A Lie group equipped with a multiplicative
Poisson bracket is called a \emph{Poisson Lie group}.

Any multiplicative Poisson bracket on $G$ identically vanishes at
its unit  element $e\in G$; its linearization at $e$ gives rise to
the structure of a Lie algebra on the dual space $\mathfrak{g}^*$;
multiplicativity then implies that the dual of the commutator map
$[\,,\,]\colon \mathfrak{g}^*\times \mathfrak{g}^*\to
\mathfrak{g}^*$ is a 1-cocycle on $\mathfrak{g}$. A pair
$(\mathfrak{g},\, \mathfrak{g}^*)$ with these properties is called a
\emph{Lie bialgebra}. A fundamental theorem, due to Drinfeld,
asserts that a multiplicative Poisson bracket on $G$ is completely
determined by its linearization and hence there is an equivalence
between the category of Poisson Lie groups (whose morphisms are Lie
group homomorphisms which are also Poisson mappings) and the
category of Lie bialgebras (whose morphisms are homomorphisms of Lie
algebras such that their duals are homomorphisms of the dual
algebras).

An action $G\times \cM\to \cM$ of a Poisson group on a Poisson
manifold $\cM$ is called a \emph{Poisson action} if this mapping is
Poisson; in other words, for  $F,H\in \mathrm{Fun}(\cM)$, their
Poisson bracket at the transformed point $g\cdot m\in \cM$ may be
computed as   follows:
\begin{equation}
\begin{aligned}
  \set{F,H}_{\cM}(g\cdot m) &= \set{\hat{F}(m,\cdot\,),\hat{H}(m,\cdot\,)}_G(g)
  + \set{\hat{F}(\cdot{}\, ,g),\hat{H}(\cdot\,
,g)}_{\cM}(m),
\end{aligned}
\end{equation}
where in the r.h.s. we set $\hat F(m, g)=F(g\cdot m)$,
 $\hat H(m,g)=H(g\cdot m)$ and treat them as functions of two variables $g\in
G$, $m\in\cM$. In that case we shall also say that the Poisson
bracket on $\cM$ is $G$-\emph{covariant}. The choice of the basic
ring of functions on $\cM$ depends on the context; we may work, for
instance, in the $C^\infty$-setting  or, alternatively, consider the
rings of polynomial or rational functions on the appropriate
manifolds.

It is sometimes useful to restrict the action $G\times \cM\to \cM$
to a subgroup of $G$. A natural class of subgroups of $G$
 are those Lie subgroups which are also Poisson submanifolds for which
the inherited Poisson structure   is of course multiplicative. This
class, however, is too restricted, since a Poisson Lie group may
have very few Poisson subgroups  and a wider class consists of the
so called \emph{admissible subgroups}. A subgroup $H\subset G$ of a
Poisson Lie group $G$ is called admissible if the subalgebra of
$H$-invariants $\mathrm{Fun}(\cM)^H\subset \mathrm{Fun}(\cM)$ is
closed with respect to the Poisson bracket. A simple admissibility
criterion is stated as follows.
 Let $\mathfrak{h}\subset \mathfrak{g}$ be the Lie  algebra
of $H$ and $\mathfrak{h}^\perp\subset \mathfrak{g}^*$ its
annihilator in $\mathfrak{g}^*$. Then $H\subset G$ is  admissible if
and only if $\mathfrak{h}^\perp\subset \mathfrak{g}^*$ is a Lie
subalgebra; $H\subset G$ is a Poisson subgroup if and only if
$\mathfrak{h}^\perp$ is an ideal in  $\mathfrak{g}^*$.

Let us assume that $H$ is admissible and that the quotient space
$\cM/H$ is smooth, so that case we may identify
$\mathrm{Fun}(\cM/H)$ with $\mathrm{Fun}(\cM)^H$ and hence the
quotient space inherits the Poisson structure. This is the basis of
\emph{Poisson reduction}, originally introduced by Lie.

The only nontrivial example which we need in the present paper is
the projective group $G=SL(2)$ (or $PSL(2)$). The group
${G}=SL(2,\mathbb{C})$ carries a family of natural Poisson
structures called the Sklyanin brackets which make
it
a Poisson Lie group. These Poisson structures are parameterized by
the choice of a classical r-matrix $r\in\mathfrak{g}\wedge
\mathfrak{g}$; for $\mathfrak{g}=\mathfrak{sl}(2)$ the classical
Yang--Baxter equation does not impose any restrictions on the choice
of $r$, so any element of $\mathfrak{g}\wedge \mathfrak{g}$ gives
rise to a Poisson bracket on $G$. It is specified by the set of
Poisson bracket relations for the matrix coefficients of $G$
(regarded as generators of its affine ring).
In usual tensor notation we have
\begin{equation}\label{Skl}
  \set{g_1,\,g_2}=[r,\,g_1\,g_2],
\end{equation}
where in the r.h.s. we regard $r\in\mathfrak{g}\wedge \mathfrak{g}$
 and $g_1\,g_2=g\otimes g$ as elements of $\Mat(2)\otimes\Mat(2)\simeq\Mat(4)$
 and compute the commutator in $\Mat(4)$.

 Let $h$, $e$, $f$ be the standard generators of $\mathfrak{sl}(2)$.
 Up to the natural equivalence there exist three types of classical
 r-matrices:
\begin{itemize}
  \item[(a)] $r=0$;
  \item [(b)] $r=h\wedge f$
  \item [(c)] $r=\epsilon \,e\wedge f$, where $\epsilon$ is a scaling
  parameter.
\end{itemize}
They correspond to three types of $G$-orbits in $\mathfrak{g}$. Case
(a) gives trivial bracket; case (c) is generic; case (b) (the so
called triangular r-matrix) is degenerate. The standard Poisson
bracket on $G$ which corresponds to case (c) is given by the
following set of relations for the matrix coefficients of
$g=\left(\begin{smallmatrix} \alpha&\beta\\
\gamma&\delta
\end{smallmatrix}\right)$, we have
\begin{equation}\label{P}
 \begin{aligned}
\set{\alpha,\beta} &= \epsilon\alpha\beta, & \set{\alpha,\gamma} &= \epsilon\alpha\gamma,\\
\set{\beta,\delta} &= \epsilon\beta\delta, & \set{\gamma,\delta} &= \epsilon\gamma\delta,\\
\set{\beta,\gamma} &=0, & \set{\alpha,\delta} &= 2\epsilon
\beta\gamma.
\end{aligned}
\end{equation}
 Notice that $\det
g=\alpha\delta-\beta\gamma$ is a Casimir function and hence the
Poisson bracket is well defined on the coordinate ring of $SL(2)$
and even of $PSL(2)$.)

In the sequel we shall be mainly concerned with the standard bracket
\eqref{P}. We shall see that the covariance condition together with
the wronskian constraint fix the Poisson structure on $G$ uniquely
up to scaling and conjugation; in particular, r-matrices of types
(a) and (b) are excluded. It will be important for us to have an
explicit description of the dual Poisson group associated with the
standard r-matrix (of type (c)\,) on $\mathfrak{g}$.

 Let $\mathfrak{b}_\pm\subset \mathfrak{g}$
be the opposite Borel subalgebras
 of $\mathfrak{g}=\mathfrak{sl}(2)$ which consist of upper (respectively, lower) triangular matrices.
 The  dual Lie algebra $\mathfrak{g}^*$ associated with the standard r-matrix may be identified with the
 subalgebra of $\mathfrak{b}_+\oplus \mathfrak{b}_-$,
\begin{equation}\label{g}
 \mathfrak{g}^* =\set{(X_+,X_-)\in \mathfrak{b}_+\oplus \mathfrak{b}_-\;|\;\diag X_++\diag X_-=0}.
\end{equation}
We conclude, in particular, that the standard Cartan subgroup $H$,
Borel subgroups $B_\pm$ and unipotent subgroups $N\pm\subset B_\pm$
are admissible subgroups of $G$.
(Of course, this is not true for conjugate
subgroups!)

The Lie group $G^*$ associated with $\mathfrak{g}^*$ may be
identified with the subgroup in $B_+\times B_-$,
\[
G^*=\set{(b_+,b_-)\in B_+\times B_-| \diag b_+\cdot \diag b_-=I}.
\]
It carries a natural Poisson bracket which makes it a Poisson Lie
group; this is the dual Poisson Lie group of $G$. The mapping
$$G^*\to G:(b_+,b_-)\mapsto M= b_+ b_-^{-1}$$ maps $G^*$ onto an open
dense subset in $G$; the induced Poisson structure on   extends
smoothly to the entire manifold $G$. Explicitly it is described by
the following formula:
\begin{equation}\label{dual}
  \{M_1,M_2\}=M_1M_2r+rM_1M_2 - M_2r_+M_1 - M_1r_-M_2,
\end{equation}
where $r_\pm=r\pm \epsilon t$ and $t\in \mathfrak{g}\otimes
\mathfrak{g}$ stands for the tensor Casimir element. This Poisson
structure on $G$ has a number of remarkable properties; in
particular, its symplectic leaves are conjugacy classes in $G$;
moreover, this bracket is covariant with respect to the action of
$G$ (equipped with the   bracket \eqref{P}) by conjugation.
Conversely, the only Poisson structure on $G$ (now regarded as a
$G$-space, not as a group) which is Poisson covariant with respect
to the action of $G$ by conjugation is that given by \eqref{dual}.

 \section{The space of wave functions as a Poisson space}

We shall assume in the sequel that all functions take values in $\mathbb{C}$.
For $M\in SL(2,\mathbb{C})$ let $\cW_M$ be the space of smooth
quasi-periodic plane curves,
\begin{equation}\label{wdef}
\cW_M=\{w\colon \mathbb{R}\to \mathbb{C}^2\vert\
w(x+2\pi)=w(x)M\,\; \text{for all}\; x\},
\end{equation}
where $w$ is denoted by a row vector. Let $\cW$ be the set of pairs,
$$\cW=\set{(w, M)|M\in SL(2,\mathbb{C}),\; w\in \cW_M }.$$ The
wronskian $W:{\cW}\rightarrow\CC$ is defined by the standard formula
\begin{equation}
W(\phi,\psi)=\phi\psi'-\phi'\psi
\end{equation}
and we define $\cW'\subset\cW $ to be the open subset consisting  of {non-degenerate curves}, i.e.
having non-zero wronskian.

We want to find the most general Poisson structure on $\cW$ which is
covariant with  respect to the right action of $G=SL(2,\mathbb{C})$
and to the action of the scaling group $\cC$. This structure appears to be partially rigid.
It is convenient to describe this Poisson structure by giving the Poisson brackets of
the `evaluation functionals' which assign to wave functions $\phi,
\psi$ their values at the running point $x\in \mathbb{R}$. The
covariance with respect to the local scaling group implies that
these brackets are quadratic and local, i.e., depend only on the
values of $\phi, \psi$ at the given points.

\begin{lemma} Assume that the Poisson bracket on $\cW$ is covariant
with  respect to the action of $\cC$. Then the Poisson structure on
$\cC$ is trivial and, writing $w=(\phi,\psi)$, the bracket of evaluation functionals has the
form
\begin{equation}\label{quad}
  \begin{aligned}
\{\phi(x),\phi(y)\}&=A(x,y)\phi(x)\phi(y), \quad \{\psi(x),\psi(y)\}=D(x,y)\psi(x)\psi(y),\\
\{\phi(x),\psi(y)\}&=B(x,y)\phi(x)\psi(y)+C(x,y)\phi(y)\psi(x).
\end{aligned}
\end{equation}
\end{lemma}
It is natural to assume that the bracket \eqref{quad} is translation
invariant, i.e., the structure functions  depend only on the
difference $x-y$. Using tensor notation, we can write these Poisson
brackets in the following condensed form:
\begin{equation}\label{exch}
  \set{w_1(x),w_2(y)}=w_1(x)w_2(y)R(x,y),
\end{equation}
where $w(x)=(\phi(x),\psi(x))$ and we write the tensor product
$w_1(x)w_2(y)$ as a row vector of length 4; the   matrix
$R(x,y)\in\Mat(4)$ is given by
\[
R(x,y)=\begin{pmatrix} A(x-y)&0&0&0\\
0&B(x-y)&-C(y-x)&0\\
0&C(x-y)&-B(y-x)&0\\
0&0&0&D(x-y)
\end{pmatrix}.
\]
Poisson brackets of this type were first studied in \cite{Babelon1}
(for a  special choice of $R$).

It is convenient to drop temporarily the Jacobi identity condition
and to consider all (generalized) Poisson brackets which are
covariant with respect to the Galois group action.
\begin{lemma} Let us assume that the Poisson bracket \eqref{exch} is
right-$G$-invariant; then the exchange matrix has the structure
\begin{equation}\label{inv}
R_0(x,y)=a(x-y)I+\begin{pmatrix} 0&0&0&0\\
0&c(x-y)&-c(x-y)&0\\
0&c(x-y)&-c(x-y)&0\\
0&0&0&0
\end{pmatrix},
\end{equation}
where $a$ and $c$ are arbitrary odd functions.
\end{lemma}
\begin{lemma}Fix  an arbitrary r-matrix $r\in \mathfrak{g}\wedge
\mathfrak{g}$ and equip $G$ with the corresponding  Sklyanin
bracket \eqref{Skl}. Let us assume that the Poisson bracket
\eqref{exch} is right-$G$-covariant; then the exchange matrix has
the structure
\begin{equation}\label{R_r}
  R_r(x,y)=R_0(x,y)+r,
\end{equation}
where we write $r\in \mathfrak{g}\wedge
\mathfrak{g}\subset\Mat(2)\otimes\Mat(2)$ as a $4\times4$-matrix in
the standard way.
\end{lemma}
For $\mathfrak{g}=\mathfrak{sl}(2)$ the classical Yang--Baxter
equation does not impose any restrictions on the choice of $r$;
indeed, it amounts to the requirement that the Schouten bracket
$[r,r]\in \mathfrak{g}\wedge \mathfrak{g}\wedge \mathfrak{g}$ should
be $\ad\, \mathfrak{g}$-invariant, but for $\mathfrak{g}=\mathfrak{sl}(2)$ we
have $\wedge^3\mathfrak{g}\simeq \mathbb{C}$. Still,  we must
distinguish two cases:
\begin{itemize}
  \item[---] $[r,r]=0$, which happens when $r=0$ or $r$ is triangular
  (cases (a) and (b) of the classification in section 2 above).
  \item[---] $[r,r]=-\epsilon^2\neq0$, which happens when $r$ is quasitriangular
  (case (c)).
  \end{itemize}
Since $R_r$ in \eqref{R_r} is the sum of 2 terms, the Schouten
bracket $[r,r]$ gives an extra term to the Jacobi identity for the
corresponding exchange bracket.
\begin{lemma}The exchange bracket \eqref{exch} with exchange matrix
\eqref{R_r} satisfies the  Jacobi identity if and only if
\begin{equation}
 \label{f0}
 \begin{aligned}
  c(x-y)c(y-z) + c(y-z)c(z-x) + c(z-x)c(x-y) &=0
  \end{aligned}
  \end{equation}
   in cases (a) and (b) and
   \begin{equation}
 \label{f}
 \begin{aligned}
   c(x-y)c(y-z) + c(y-z)c(z-x) + c(z-x)c(x-y) &=-\epsilon^2
  \end{aligned}
\end{equation}
in case  (c).
\end{lemma}
Functional equation \eqref{f} is a version of the so called
Rota--Baxter equation. To solve it, one can put $c(x)=\epsilon\,
C(x)$ and express $C$ as a Cayley transform,
 $$C(x)=\frac{f(x)+1}{f(x)-1};$$
 then \eqref{f}
immediately yields for $f$ the standard 2-cocycle relation
\[
f(x-y)f(y-z)f(z-x)=1.
\]
The obvious solution is thus
  $C_\lambda(x-y)=\coth\lambda(x-y)$,
where $\lambda$ is a parameter. Setting $\lambda\to\infty$, we
obtain a particular solution $C(x-y)=\sign(x-y)$. We shall see that
this special solution is the only one which is compatible with the
 constraint $W=1$. The
solution of the degenerate equation \eqref{f0} is $c(x)=1/x$.

So far, the most general Poisson structure on $\cW$ still contains
functional moduli and a free parameter. As is easy to check, the
  Poisson brackets for the ratio $\eta=\phi/\psi$ do not
depend on $a$:
\begin{proposition} We have
\begin{equation}\label{etapb}
\{\eta(x),\eta(y)\}=\epsilon\left(\eta(x)^2-\eta(y)^2\right)-c(x-y)\left(\eta(x)-\eta(y)\right)^2.
\end{equation}
\end{proposition}
\begin{remark}
Formula \eqref{etapb} defines a family of $G$-covariant Poisson
brackets on the space of projective curves. However, in order to
establish a connection between these brackets and   Schroedinger
operators we must take into account the wronskian constraint which
restricts the choice of $c$. The second structure function $a$ drops
out after projectivization and is not restricted by the Jacobi
identity. We shall see, however, that the wronskian constraint
suggests a natural way to   choose $a$ as well. An interpretation of
the general family \eqref{etapb} of Poisson brackets remains an open
question.
\end{remark}

Our next proposition describes the basic Poisson bracket relations
for the wronskian:
\begin{proposition} We have
  \begin{multline}\label{w}
  \set{W(x),\phi(y)}=(c(x-y)-2a(x,y))W(x)\phi(y)\\-c'(x-y)\phi(x)[\phi(x)\psi(y)-\psi(x)\phi(y)].
  \end{multline}
  By symmetry, a similar formula holds for $\set{W(x),\psi(y)}$.
\end{proposition}
Formula \eqref{w} immediately leads to the following crucial
observation:
\begin{proposition}\label{main}
The constraint $W=1$ is compatible with the Poisson brackets for
scaling invariant $\eta$ if and only if the last term in \eqref{w}
is identically zero; this is possible if and only if
$C'(x-y)$ is a multiple of $\delta(x-y)$, i.e., if
$C(x-y)$ is a multiple of $\sign(x-y)$.
\end{proposition}
  It is important
that the wronskian constraint excludes the possibility that
$\epsilon=0$  and hence the corresponding Poisson structure on $G$
is conjugate to the standard one (case (c)). From now on, 
without restricting the generality, we fix $\epsilon=1$.
\begin{proposition} Let us assume that $c(x-y)=\sign(x-y)$; then the
Poisson bracket relations for the wronskian are given by:
\begin{equation}\label{pbw}
    \{W(x),W(y)\}=(\sign(x-y)-2a(x,y))W(x)W(y),
 \end{equation}
or, equivalently
\begin{equation}\label{pbw-log}
   \{\log W(x),\log W(y)\}=(\sign(x-y)-2a(x,y)).
\end{equation}
\end{proposition}
Formulae \eqref{w} and \eqref{pbw} suggest the following
distinguished choice of $a$:
\begin{proposition}
Assume that $a$ is so chosen that
$$\sign(x-y)-2a(x,y)=\delta'(x-y).$$  (In other words, $a(x,y)$ is the distribution kernel of the operator
 $\frac{1}{2}\left(\d^{-1}-\d\right)$.) Then: (i) The logarithms of
wronskians form a Heisenberg Lie algebra, the central extension of
the abelian Lie algebra of $\cC$. (ii) Let $\cC'=\cC/\mathbb{C}^*$
be the quotient of the scaling group over the subgroup of constants;
$\log W$ is the moment map for the action of $\cC'$ on $\cW$.
\end{proposition}
Recall that according to the general theory the Poisson bracket
relations for the moment map may reproduce the commutation relations
for a central extension of the original Lie algebra. This is
precisely what happens in the present case.

With this choice of $a$ and $C$ the Poisson geometry of the space
$\cV$ of wave functions becomes finally quite transparent: $\cV$
 arises as a result of Hamiltonian
reduction with respect to $\cC$  over the zero level of the
associated moment map. The constraint set $\log W=0$ is (almost)
non-degenerate (i.e., this is a 2$^\text{nd}$ class constraint,
according to Dirac).
 The projective invariants commute with the
wronskian and hence their Poisson brackets are not affected by the
constraint.\footnote{With this choice of $a$ the Poisson structure
on $\cW$ becomes non-degenerate; for other possible choices this may
be not true. For example, the opposite possibility is to set
$\sign(x-y)-2a(x,y)=0$. This makes the bracket on $\cW$ highly
degenerate; its kernel is eliminated by the wronskian constraint,
and the reduced structure remains the same. While logically
possible, the resulting picture is much less attractive. }

The description of the Poisson structure on $\cV$ is completed by
the Poisson brackets for the monodromy.
\begin{proposition} The Poisson covariant brackets for the monodromy
have the form
\begin{equation}\label{Mb}
\begin{aligned}
\{w(x)_1,M_2\}&=w(x)_1\bigl[M_2r_+ - r_-M_2\bigr],\\
\{M_1,M_2\}&=M_1M_2r+rM_1M_2 - M_2r_+M_1 - M_1r_-M_2.
\end{aligned}
\end{equation}
\end{proposition}
 The Poisson
bracket for the monodromy is precisely the Poisson bracket of the
dual group $G^*$ described in \eqref{dual}. In other words, the
`forgetting map' $\mu: (w, M)\mapsto M$ is a Poisson morphism from
$\cW$ into the dual group $G^*$.\footnote{The Poisson bracket
\eqref{dual} is  ubiquitous in various problems related to
monodromy; another striking example, which is very close to our
present context, is its r\^ole in the theory of isomonodromic
deformations described in the very interesting paper of P.Boalch
\cite{Boalch}.} This mapping is of special importance.

\begin{proposition}
The mapping $\mu$ is the non-abelian moment map\footnote{We refer
the reader for instance to \cite{Babelon} for the general definition
of non-abelian moment maps associated with Poisson group actions.}
associated with the right action of $G$ on $\cW$.
\end{proposition}

Let us now list the Poisson bracket relations in the differential
algebra $\mathbb{C}\langle\eta\rangle$ and its various subalgebras
which correspond to different admissible subgroups of $G$.

\begin{proposition}
(i) Consider the tower of differential extensions
\begin{equation*}\label{tower}
 \xymatrix{ &&\mathbb{C}\langle\eta\rangle&&\\
&\mathbb{C}\langle\eta\rangle^H\ar@{^{(}->}[ur]&&\mathbb{C}\langle\eta\rangle^N\ar@{^{(}->}[ul]\\
&&\mathbb{C}\langle\eta\rangle^B\ar@{^{(}->}[ur]\ar@{^{(}->}[ul]\ar@{^{(}->}[uu]&&\\
 &&\mathbb{C}\langle\eta\rangle^G\ar@{^{(}->}[u]}
\end{equation*}
All arrows in this commutative diagram are Poisson morphisms.

(ii) The basic Poisson bracket relations in
$\mathbb{C}\langle\eta\rangle$ are given by
\begin{equation}\label{eta-final}
  \{\eta(x),\eta(y)\}=\eta(x)^2-\eta(y)^2-\sign(x-y)\left(\eta(x)-\eta(y)\right)^2.
\end{equation}

(iii) We have $\mathbb{C}\langle\eta\rangle^N\simeq
\mathbb{C}\langle\theta\rangle$, where
  $\theta:=\eta'$; moreover,
\begin{equation}\label{theta}
\{\theta(x),\theta(y)\}=
2\sign(x-y)\theta(x)\theta(y).
\end{equation}

(iv)  The subalgebra of $B$-invariants  is generated by
$v:=\half\eta''/\eta'=\half\theta'/\theta$; we have:
\begin{equation}\label{v}
 \{v(x),v(y)\}=\tfrac12\delta'(x-y).
\end{equation}
(v)
 The subalgebra of $G$-invariants  is generated by $u=\half
S(\eta)=v'-v^2$; we have:
\begin{equation}\label{u}
 \{u(x),u(y)\}=\tfrac12\delta'''(x-y) + \delta'(x-y)\bigl[u(x)+u(y)\bigr].
\end{equation}
\end{proposition}
Formula \eqref{u} reproduces the standard Virasoro algebra; in other
words, the Poisson algebra \eqref{eta-final} constructed from
general covariance principles is indeed an extension of the
Poisson--Virasoro algebra.
\begin{remark}
The Poisson bracket relations \eqref{theta} -- \eqref{u} listed above are particularly simple, since their r.h.s. is algebraic.
 Because the basic Poisson bracket relations \eqref{eta-final} are nonlocal, this  need not always be the case. This is what happens in the case of $H$-invariants:
\end{remark}
\begin{proposition}(i) The differential subalgebra of $H$-invariants in $\mathbb{C}\langle\eta\rangle$ is generated
by $\rho=\eta'/\eta$.
(ii) The  Poisson brackets for $\rho$ have the form
$$
\set{\rho(x),\,\rho(y)}=2\rho(x)\rho(y)\left[\sinh\int_x^y\rho(s)\,ds+\sign(x-y)\cosh\int_x^y\rho(s)\,ds\right].
$$
\end{proposition}

It is well known that the standard KdV
 equation is generated with respect to the Virasoro bracket
by  the Hamiltonian
\begin{equation}\label{kdv}
  H=\int u^2\,dx.
\end{equation}
The    Hamiltonians of all higher KdV equations are associated with
trace identities for $H_u$ and hence are $G$-invariant; they
generate a system of compatible commuting flows on all levels of the
extension tower.

\begin{proposition}
The following commutative diagram which is formed by Poisson maps summarizes all information
on the evolution equations generated by the standard Hamiltonian
\eqref{kdv} and on the differential substitutions which relate these
equations.

{\footnotesize

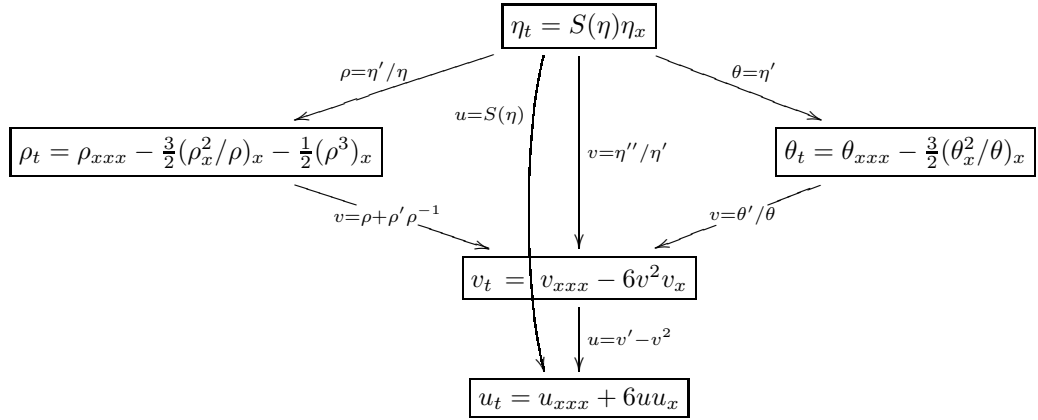
$$
\xymatrix{&&\text{\fbox{$\eta_t=S(\eta)\eta_x$}}\ar@(ul,dl)[ddd]_(.35){u=S(\eta)}\ar@{->}[dd]^{v=\eta''/\eta'}\ar@{->}[dl]_{\rho=\eta'/\eta}\ar@{->}[dr]^{\theta=\eta'}&&\\
&\text{\fbox{$\rho_t=\rho_{xxx}-\tfrac32(\rho_x^2/\rho)_x-\tfrac12(\rho^3)_x$}}\ar@{->}[dr]|{v=\rho+\rho'\rho^{-1}}&&\text{\fbox{$\theta_t=\theta_{xxx}-\tfrac32(\theta_x^2/\theta)_x$}}\ar@{->}[dl]|{v=\theta'/\theta}\\
&&\text{\fbox{$v_t\,=\;v_{xxx}-6v^2v_x$}}\ar@{->}[d]^{u=v'-v^2}&&\\
&&\text{\fbox{$u_t=u_{xxx}+6uu_x$}}&&
}
$$
}
\end{proposition}
 All equations in this diagram belong to the well known class of ``equations of the KdV
 type''. Their mutual relations were
  discussed by Wilson
\cite{wilson1}, although the  Hamiltonian description which we
propose is totally different. Equation
\begin{equation}\label{urKdV}
  \eta_t=S(\eta)\eta_x=\eta_{xxx}-\tfrac32\eta_{xx}^2/\eta_x,
\end{equation}
 is sometimes called the Schwarz--KdV equation; in
\cite{wilson1} George Wilson suggested for it the name  ``ur-KdV
equation'', due to its position atop the extension tower.
\begin{remark}Equations which appear in the diagram form a rather
small part in the general class of ``equations of the KdV
 type'' discussed in \cite{SS}, where a classification theorem is
 given for evolution equations of the form $u_t=u_{xxx}+F(u, u_x,
 u_{xx})$ which admit nontrivial conservation laws. General
 equations  of this type   depend on several parameters and may
 include elliptic functions, as was first noticed by Calogero and
 Degasperis \cite{CD}. We expect that rational equations from this
 list will also fit into the  Poisson group setting by bringing into
 play the wave functions for different values of energy, as
 suggested in \cite{ds1}.
\end{remark}
\section{Discrete case}
The theory of the Schroedinger equation has a simple and natural
lattice counterpart. Consider the 2$^\text{nd}$ order difference
equation on the one-dimensional lattice with   periodic potential
\begin{equation}\label{dhill}
  \phi_{n+2}+u_n\phi_{n+1}+\phi_{n}=0, \q u_{n+N}=u_n.
\end{equation}
Let $\tau$ be the shift operator, $(\tau\phi)_n=\phi_{n+1}$.
Equation \eqref{dhill} may be written in operator form as
\begin{equation}\label{dhill_op}
  \left(\tau^2+u\,\tau+1\right)\phi=0.
\end{equation}
For a given $u$, the space of its solutions is two-dimensional; any
two solutions $\phi, \psi$ have constant wronskian
$W=\phi_n\psi_{n-1}-\phi_{n-1}\psi_n$.  The monodromy matrix $M$ is
defined in the standard way.

 The projective description of discrete
  Schroedinger equations is given by the following theorem.  To
state it we need a few elementary notions. An ordered projective
configuration is a map $\gamma: \mathbb{Z}\to \mathbb{C}P_1$;  we
shall simply speak of projective configurations, for short. A
configuration is called non-degenerate if $\gamma_n\neq
\gamma_{n+1}$. for all $n$. A plane configuration is a map
$w:\mathbb{Z}\to \mathbb{C}^2$; it is called non-degenerate if
$w_n\wedge w_{n+1}\neq0$. We denote $w_n$ by the row vector
$(\phi_n, \psi_n)$.
\begin{thm}
(i) Any pair of linearly independent solutions of the discrete
Schroedinger equation defines a  non-degenerate quasi-periodic
projective configuration $\gamma:\mathbb{Z}\to \mathbb{C}P_1$ such
that $\gamma_{n+N}=\gamma_n\cdot M$. Any two projective
configurations associated with a given discrete   Schroedinger
equation are related by a global projective transformation. (ii)
Conversely, any non-degenerate quasi-periodic projective
configuration may be lifted to a non-degenerate plane configuration
such that its wronskian is equal to 1.
\end{thm}
As before, we replace the projective line with its affine model
putting $\eta_n=\phi_n/\psi_n$. The   group $G=SL(2)$ is the
(difference) Galois group of equation \eqref{dhill}.  Curiously, the
potential $u$ itself is \emph{not} a rational Galois invariant. A
natural finite difference analog of the Schwarzian derivative is the
 cross-ratio,
$$
s_n[\eta]:=[\eta_{n},\eta_{n+1},\eta_{n+2},\eta_{n+3}]=\frac{\eta_n-\eta_{n+2}}{\eta_n-\eta_{n+1}}\cdot\frac{\eta_{n+1}-\eta_{n+3}}{\eta_{n+2}-\eta_{n+3}};
$$
an elementary calculation yields
\begin{equation}\label{anh}
  s_n=u_nu_{n+1}.
\end{equation}
From now on  we shall assume that the period $N$ of the lattice is
\emph{odd}. In this  case
  the potential   may be restored as the periodic solution of
\eqref{anh} (regarded as an equation for $u$ for given
$s_m[\eta]\left|_{m=1}^N\right.$); it  belongs to a quadratic
extension of $\mathbb{C}(\eta)^G=\mathbb{C}(
u\,u^\tau)\subset\mathbb{C}( u)$. Note that the resulting formula is
non-local, that is, it depends on the values of $\eta_m$ for all
$m$.

The Poisson structure on the space of discrete   Schroedinger
operators is much less obvious than in the continuous case; it may
be regarded as a lattice analog of the Virasoro algebra. One version
of its definition was proposed in \cite{frs} as a part of a more
general theory, the q-difference version of the Drinfeld--Sokolov
theory \cite{ds} which applies to q-difference equations of
arbitrary order (see also \cite{ss}). Another definition of the
lattice Virasoro algebra had been proposed earlier by Faddeev and
Takhtajan
 \cite{ft}.  The
projective point of view outlined in the present paper  also yields
a natural Poisson structure on the space of discrete Hill's
operators; we shall see that it is identical to that introduced in
\cite{frs} and is simply related to the Faddeev--Takhtajan bracket.

In this section we shall denote by $\cW$ the space of all plane
quasi-periodic configurations and by $\cC$ the discrete scaling
group.
\begin{proposition} (i) Let us assume that the Poisson structure on
$\cW$ is covariant with respect to the right action of $G$ and to
the natural action of the scaling group. Then the bracket between
the evaluation functionals is given by
\begin{equation}\label{exch-lat}
  \set{w^1_m(x),w^2_n(y)}=w^1_m(x)w^2_n(y)R(m-n),
\end{equation}
where
\begin{equation}\label{exch-d}
   R(k)=R_0(k)+r,\q R_0(k)=a_kI+\begin{pmatrix} 0&0&0&0\\
0&c_k&-c_k&0\\
0&c_k&-c_k&0\\
0&0&0&0
\end{pmatrix},
\end{equation}
(we omitted Poisson bracket relations for the monodromy which remain
the same as before). Here $a_{k}$ is an arbitrary odd function and
$c_{k}$ is an odd function which satisfies
\begin{equation}\label{f-discr}
  c_{n-m}c_{m-k} + c_{m-k}c_{k-n} + c_{k-n}c_{n-m} =\alpha,
\end{equation}
where $\alpha=0$ when $r$ is a trivial or triangular r-matrix and
$\alpha=-\epsilon^2$ for $r$ quasitriangular (case (c)).
\end{proposition}

The wronskian $W$ of a plane configuration $w=(\phi, \psi)$ is
defined by the obvious formula
$$W[w]_n=\phi_n\psi_{n-1}-\psi_n\phi_{n-1}.$$
The space $\cV\subset\cW$ of wave functions of discrete
Schroedinger operators is defined by the constraint $W[w]=1$.
\begin{proposition} We have
  \begin{multline}\label{w-discr}
  \set{W_n,\phi_m}=(a_{n-m}+a_{n-1-m}-c_{n-m})W_n\phi_m\\+(c_{n-m}-c_{n-m-1})(\phi_n \phi_{n-1}\psi_m-\phi_n\psi_{n-1}\phi_m).
  \end{multline}
A similar formula holds for $\set{W_n,\psi_m}$.
\end{proposition}
Scaling invariants $\eta_n$ commute with the wronskian if and only
if the second term in \eqref{w-discr} is also proportional to
$W_n\phi_m$; this condition implies that
\begin{equation}\label{cond_discr}
 c_{n-m}-c_{n-m-1}=\epsilon(\delta_{nm}+\delta_{n,m+1}).
\end{equation}
 Without restricting the generality we may assume that
$\epsilon=1$ and in that case we get
\begin{equation}\label{w-discr_good}
 \set{W_n,\phi_m}=(a_{n-m}+a_{n-1-m}-c_{n-m}+\delta_{nm}+\delta_{n,m+1})W_n\phi_m.
\end{equation}
Fortunately,
 condition \eqref{cond_discr}  is again satisfied by the sign
function  and hence the
  Poisson structure on the  space of projective configurations remains basically the
  same as in the continuous case. Moreover, if $\eta$ is a
  projective curve, which defines a  Schroedinger equation, we may fix a
  generic
  set of values $\set{x_1,\dots,x_N}$ of the coordinate $x$ on the
  circle such that $\eta(x_n)\neq \eta(x_{n+1})$; then   $\set{\eta(x_n)}$ is a
  non-degenerate projective configuration which gives rise to a
  difference   Schroedinger equation and the evaluation functionals $\eta\mapsto\eta(x_n)$
  form a Poisson subalgebra in the big Poisson algebra
  \eqref{eta-final}.  Explicitly we have
\begin{equation}\label{eta-discr}
     \{\eta_n,\eta_m\}=\eta_n^2-\eta_m^2-\sign(n-m)\bigl(\eta_n-\eta_m\bigr)^2.
\end{equation}
Note that it's of course not true
  that the \emph{solutions} of this difference equation are the values of
  the wave functions for the continuous equation: indeed, the
  wronskian constraints are \emph{different} in the two cases. It is
 noteworthy that nevertheless the conditions imposed by these
  constraints on the structure function $c$ are satisfied by the
  same standard function.

In order to compute the Poisson structure induced by
 \eqref{eta-discr} on the set of potentials let us
start with the subfields of rational $N$- and $B$-invariants in
$\mathbb{C}(\eta)$; in complete analogy with the continuous case we
have $\mathbb{C}(\eta)^N=\mathbb{C}(\theta)$, where
$\theta_m:=\eta_{m+1}-\eta_m$, and
$\mathbb{C}(\eta)^B=\mathbb{C}(\lambda)$, where
$$
\lambda_m:=\frac{\eta_{m+2}-\eta_{m+1}}{\eta_{m+1}-\eta_m}\cdot
$$
An easy computation yields
\begin{equation}\label{lambda}
 \{\theta_m,\theta_n\}= -2\sign(m-n)\theta_m\theta_n,\q
\{\lambda_m,\lambda_n\}= 2\bigl(\delta_{m+1,n} - \delta_{m,n+1}\bigr)\lambda_m\lambda_n.
\end{equation}
A natural interpretation of the variables $\lambda_n$ is connected
with the Miura transform for the  discrete   Schroedinger equation.
Let us assume that the difference operator \eqref{dhill_op} is
factorized,
\begin{equation}\label{hill_factor}
 \tau^2+u\,\tau+1 = (\tau+v)(\tau+v^{-1}).
\end{equation}
The potentials $u$, $v$ are related by the difference Miura map,
\begin{equation}\label{miura}
 u_n=v_n+v_{n+1}^{-1}.
\end{equation}
We may assume without restricting the generality that $\psi$ is the
solution of \eqref{dhill} which satisfies the first order equation
$(\tau+v^{-1})\psi=0$.    Let $\phi$ be the second solution of this
equation such that $W(\phi, \psi)=1$ and $\eta=\phi/\psi$; then
$$\eta_{n+1}-\eta_n=\frac{1}{\psi_n\psi_{n+1}}.$$  Clearly, $
v_n=-{\psi_{n}}/{\psi_{n+1}}$ and hence
\begin{equation}\label{vv}
 v_nv_{n+1}=\frac{\psi_n}{\psi_{n+2}}
=\frac{\psi_{n+1}\psi_{n}}{\psi_{n+2}\psi_{n+1}}=\frac{\eta_{n+2}-\eta_{n+1}}{\eta_{n+1}-\eta_{n}} =
\lambda_n
\end{equation}
 Thus
 $\lambda_n$ is the product of two neighbouring potentials in the
factorized   Schroedinger operator \eqref{hill_factor}. The
potentials themselves again are not rational Galois invariants of
$B$ and belong to a quadratic extension of $\mathbb{C}(\lambda)$.
From \eqref{miura}, \eqref{vv} we easily derive that
\begin{equation}\label{s_n}
 s_n=u_nu_{n+1}=\frac{(1+\lambda_n)(1+\lambda_{n+1})}{\lambda_{n+1}}.
\end{equation}
\begin{proposition}
We have
\begin{equation}
\begin{aligned}\label{mean}
\set{\lambda_m,\lambda_n}&=(\delta_{m+1,n}-\delta_{m,n+1})\lambda_m\lambda_n,\\
\{s_m,s_n\}&=\bigl(\delta_{m+1,n}-\delta_{m,n+1}\bigr)(s_m+s_n-s_ms_n)\\
&\quad+ s_ms_n\bigl(s_{m+1}^{-1}\delta_{m+2,n} -
s_{n+1}^{-1}\delta_{m,n+2}\bigr).
\end{aligned}
\end{equation}
\end{proposition}
Formula \eqref{mean} implies the following Poisson bracket relations
for the potentials:
\begin{proposition}
Let $\Phi_n=(-1)^n\sign n, \; n\neq 0,\; \Phi_0=0 $. Then
\begin{equation}\label{lattice vir}
  \set{v_n,v_m}=2\Phi_{n-m}v_nv_m  \; \text{and}\;   \set{u_n,u_m}=2\Phi_{n-m}u_nu_m+2(\delta_{m+1,n}-\delta_{m,n+1}).
\end{equation}
\end{proposition}
%
%
Formula \eqref{lattice vir} coincides with the lattice Virasoro
algebra introduced in \cite{frs}, while \eqref{mean}  coincides with
the Faddeev--Takhtajan version of the lattice Virasoro algebra. The
non-locality of the Poisson bracket relations in \eqref{lattice vir}
is due to the non-locality of the formula for potentials $v$ and $u$
in terms of $\eta$. The same structure constants $\Phi_{n-m}$ arise
in \cite{frs} in the framework of the discrete Drinfeld--Sokolov
theory, which provides for this formula a totally different  (and
more direct) explanation.

\section*{Acknowledgement}
 The authors would like to thank L.D.Faddeev,  V.Fock and   V.Sokolov for useful
 discussions. The work of the second author was partially supported
 by the INTAS-OPEN grant 03-51-3350, the RFFI grant 05-01-00922  and
 the  ANR program ``GIMP'' ANR-05-BLAN-0029-01. The first author is
 grateful to the Association Suisse-Russe for financing his visit to
 the Steklov Institute, with special thanks to J.-P.Periat and S.Yu.Sergueeva.

\end{document}